\documentclass{raa}
\usepackage{graphicx,times}
\usepackage{amssymb,amsmath}
\usepackage{natbib}
\bibpunct{(}{)}{;}{a}{}{,}

\begin{document}

\title{Determining proportions of lunar crater populations by fitting crater size distribution}

\volnopage{Vol.0 (200x) No.0, 000--000}      
\setcounter{page}{1}          

\author{Nan Wang
	\inst{}
\and Ji-Lin Zhou
	\inst{}}
	
\institute{School of Astronomy and Space Science and Key Laboratory of Modern Astronomy and Astrophysics in Ministry of Education, Nanjing University, Nanjing 210046, China; {\it zhoujl@nju.edu.cn}}

\date{Received~~2009 month day; accepted~~2009~~month day}

\abstract{
We determine the proportions of two mixed crater populations distinguishable by size distributions on the Moon. A "multiple power-law" model is built to formulate crater size distribution $N(D) \propto D^{-\alpha}$ whose slope $\alpha$ varies with crater diameter $D$. Fitted size distribution of lunar highland craters characterized by $\alpha = 1.17 \pm 0.04$, $1.88 \pm 0.07$, $3.17 \pm 0.10$ and $1.40 \pm 0.15$ for consecutive $D$ intervals divided by 49, 120 and 251\,km and that of lunar Class 1 craters with a single slope $\alpha = 1.96 \pm 0.14$, are taken as Population 1 and 2 crater size distribution respectively, whose sum is then fitted to the size distribution of global lunar craters with $D$ between 10 and 100\,km. Estimated crater densities of Population 1 and 2 are $44 \times 10^{-5}$ and $5 \times 10^{-5}$\,km$^{-2}$ respectively, leading to the proportion of the latter $10 \%$. The results underlines the need for considering the Population 1 craters and the relevant impactors, the primordial main-belt asteroids which dominated the late heavy bombardment.
}

\authorrunning{N. Wang \& J.-L. Zhou}
\titlerunning{Proportions of lunar crater populations}

\maketitle

\section{Introduction}\label{sec-intro}

The crater record of the Moon and terrestrial planets are helpful in understanding the evolution of solar system. Size distribution of craters has been used to infer the properties of impactors that generated them.
\cite{Strom2005, Strom2015} examined crater size distributions of various regions on the Moon, Mars, Venus and Mercury and identified two crater populations related to two impactor populations in the inner solar system. Population 1 craters with wavy size distributions were found on heavily cratered surfaces, while Population 2 craters with smooth size distributions were primarily found on lightly cratered and younger plains.
After the size distributions of impactors were derived from those of craters, the apparent matches were shown between the Population 1 impactors and the contemporary main-belt asteroids (MBAs) and between the Population 2 impactors and the near-Earth objects (NEOs). Since the main-belt size distribution changed little after the first $\sim 100$\,Myr (\citealt{Bottke2005}), \cite{Strom2005, Strom2015} suggested the primordial MBAs and the current NEOs were the Population 1 and 2 impactors, respectively.
They further indicated that the former impactor population dominated during the late heavy
bombardment (LHB), a sudden planetesimal bombardment to the inner solar system triggered by the migration of giant planets $\sim 4.1$\,Gya (\citealt{Morbidelli2012, Marchi2013b, Marchi2013a}), while the latter followed and dominated until present, being constantly resupplied from the main belt by Yarkovsky and YORP effects.
\cite{Marchi2009} supported those suggestions by computing the lunar crater size distribution based on modeled impactor flux and comparing the computation to the observation. The crater size distributions of the highlands and the Nectaris Basin, the oldest regions on the Moon, were found to be best fitted with the MBAs being impactors.
\cite{Head2010} also confirmed the difference in size distribution between two lunar crater populations on the pre- and post-mare regions, which are older and younger than the Orientale Basin respectively.


What should be mentioned is that \cite{Marchi2012} implied the possibility of another impactor population dominating the pre-LHB epoch, which is called Population 0 here, having half the mean impact speed of Population 1 as well as the size distribution of the main belt. Still, this work is based on the assumption that the major lunar impactors are Population 1 and 2, following \cite{Strom2015}.

Different crater populations probably have different cratering distributions. At present, there have been more sufficient studies on lunar cratering by Population 2 impactors (\citealt{Gallant2009, Ito2010, LeFeuvre2011, Kawamura2011, Oberst2012}) than Population 1 (\citealt{Wang2016}), and the revised cratering chronology method have been usually based on the cratering asymmetry of the current impactors only (\citealt{Morota2005, LeFeuvre2011}). However, according to \cite{Strom2015}, for craters larger than 10\,km, the density of Population 1 exceeds that of Population 2 by more than one order of magnitude. That emphasizes the importance of the former population. We will further examine this point by quantitatively determine the proportions of two crater populations of the Moon.
In Section \ref{sec-methods&results}, using our "multiple power-law" model, the size distribution of global lunar craters is fitted, resulting in the fitted amounts of two crater populations mixed on lunar surface. Section \ref{sec-discussion} presents discussion about cratering asymmetry and cratering chronology, and Section \ref{sec-conclusion} is our conclusion.

\section{Methods and results}		 \label{sec-methods&results}

\subsection{Multiple Power-law Model}    \label{sec-MPL model}

The cumulative crater size distribution is commonly assumed to be $N(D) \propto D^{-\alpha}$, where $D$ is crater diameter, $N(D)$ is number of craters with diameters larger than $D$ per unit area and $\alpha$ is the power-law slope. We here propose a "multiple power-law" model, which is inspired by the "broken power law" model of \cite{Ivezic2001}, to formulate the crater size distributions with complex shapes.

The model assumes a crater size distribution can be divided into $n$ parts by limits $D_{0,1,\dots,n}$ so that in each $D$ interval $[D_i,D_{i+1}]$ the relevant $\alpha_{i}$ is invariant ($i = 0,1,\dots,n-1$). Thus, the cumulative size distribution (CSD) $N(D)$, differential size distribution (DSD) $N'(D) = |{\rm{d}}N/{\rm{d}}D|$ and relative size distribution (RSD) $R(D) = N'(D) D^3$ are described as
\begin{align}
  N(D)  &= C_i D^{-\alpha_i} + I_i,       \quad (D_i \le D \le D_{i+1})       \label{eq-CSD}   \\
  N'(D) &= \alpha_i C_i D^{-\alpha_i-1},  \quad (D_i \le D \le D_{i+1})       \label{eq-DSD}   \\
  R(D)  &= \alpha_i C_i D^{-\alpha_i+2}.  \quad (D_i \le D \le D_{i+1})       \label{eq-RSD}
\end{align}
Expressions of coefficients $C_{0,1,\dots,n-1}$ and $I_{0,1,\dots,n-1}$ are derived as follows.

When $D=D_i$ and $i \ne 0$, Equation \ref{eq-DSD} leads to
\begin{align}
  &N'(D_i)= \alpha_i C_i D_i^{-\alpha_i-1} = \alpha_{i-1} C_{i-1} D_i^{-\alpha_{i-1}-1},         \\
  \Rightarrow \quad & \frac{C_i}{C_{i-1}} = \frac{\alpha_{i-1}}{\alpha_i} D_i^{\alpha_i-\alpha_{i-1}},     \\
  \Rightarrow \quad & \frac{C_i}{C_0} = \prod\limits_{j=1}^{i}\frac{C_j}{C_{j-1}} = \frac{\alpha_{0}}{\alpha_i} \prod\limits_{j=1}^{i}D_j^{\alpha_j-\alpha_{j-1}}.
\end{align}
Thus, $C_{0,1,\dots,n-1}$ are expressed to be
\begin{equation}    \label{eq-C_i}
  C_i =
  \begin{cases}
    C_0, & (i=0) \\
    C_0 \frac{\alpha_0}{\alpha_i} \prod\limits_{j=1}^{i}D_j^{\alpha_j-\alpha_{j-1}}. & (i = 1,2,\dots,n-1)
  \end{cases}
\end{equation}
When $D=D_n$, Equation \ref{eq-CSD} leads to
\begin{align}
  & N(D_n) = C_{n-1} D_n^{-\alpha_{n-1}} + I_{n-1},   \\
  \Rightarrow \quad & I_{n-1} = N(D_n) - C_{n-1} D_n^{-\alpha_{n-1}}, \label{eq-I_n-1}
\end{align}
while when $D=D_{i+1}$ and $i \ne n-1$,
\begin{align}
  & N(D_{i+1}) = C_i D_{i+1}^{-\alpha_i} + I_i = C_{i+1} D_{i+1}^{-\alpha_{i+1}} + I_{i+1},  \\
  \Rightarrow \quad & I_i - I_{i+1} = C_{i+1}D_{i+1}^{-\alpha_{i+1}} - C_i D_{i+1}^{-\alpha_i},     \\
  \Rightarrow \quad & I_i - I_{n-1} = \sum\limits_{j=i}^{n-2}(I_j - I_{j+1})  \\
   = & -C_i D_{i+1}^{-\alpha_i} + \sum\limits_{j=i+1}^{n-1}C_j(D_j^{-\alpha_j}-D_{j+1}^{-\alpha_j}) + C_{n-1} D_n^{-\alpha_{n-1}}.    \label{eq-I_i-I_n-1}
\end{align}
Substituting Equation \ref{eq-I_n-1} into \ref{eq-I_i-I_n-1}, it turns out
\begin{equation}  \label{eq-I_i}
  I_i = -C_i D_{i+1}^{-\alpha_i} + G_{i+1} + N(D_n),    \quad (i = 0,1,\dots,n-1)
\end{equation}
where
\begin{equation}  \label{eq-G_i}
  G_i =
  \begin{cases}
    \sum\limits_{j=i}^{n-1}C_j(D_j^{-\alpha_j}-D_{j+1}^{-\alpha_j}),    & (i=0,1,\ldots,n-1)  \\
    0 .                                                                 & (i=n)
  \end{cases}
\end{equation}

Rewriting Equation \ref{eq-CSD} with \ref{eq-I_i}, we derive the general formulation of CSD
\begin{equation}    \label{eq-CSD2}
  N(D) = C_i(D^{-\alpha_i}-D_{i+1}^{-\alpha_i}) + G_{i+1} + N(D_n). \quad (D_i \le D \le D_{i+1})
\end{equation}
Coefficients $C_{0,1,\dots,n-1}$ and $G_{0,1,\dots,n}$ (Eq. \ref{eq-C_i} and \ref{eq-G_i}) are totally determined by power-law slopes $\alpha_{0,1,\ldots,n-1}$ and interval limits $D_{0,1,\ldots,n}$ as well as $C_0$.

Furthermore, we show that $C_0$ will vanish when $D_n = +\infty$ and $N(D)$ is normalized. The normalized CSD is defined as
\begin{equation}    \label{eq-norCSD}
  \bar{N}(D) = \frac{N(D)}{N(D_0)},
\end{equation}
where $N(D_0) = G_0 + N(D_n)$. If $D_n = +\infty$, i.e., $N(D_n)=0$, then
\begin{equation}    \label{eq-norCSD2}
  \bar{N}(D)= \frac{\bar{C}_i(D^{-\alpha_i}-D^{-\alpha_i}_{i+1})+\bar{G}_{i+1}} {\bar{G}_0},   \quad (D_i \le D \le D_{i+1})
\end{equation}
where
\begin{align}
  &\bar{G}_i =
  \begin{cases}
    \sum\limits_{j=i}^{n-1}\bar{C}_j(D_j^{-\alpha_j}-D_{j+1}^{-\alpha_j}),    & (i=0,1,\ldots,n-1)  \\
    0,                                                                              & (i=n)
  \end{cases}   \\
  &\bar{C}_i=
  \begin{cases}
    1,                                                                               & (i=0)     \\
    \frac{\alpha_0}{\alpha_i} \prod\limits_{j=1}^{i}D_j^{\alpha_j-\alpha_{j-1}}.     & (i=1,2,\ldots,n-1)
  \end{cases}
\end{align}
Thus, once $\alpha_{0,1,\ldots,n-1}$ and $D_{0,1,\ldots,n}$ are given, $\bar{N}(D)$ can be obtained directly.

The general formulations of CSD, DSD and RSD (Eq. \ref{eq-CSD2}, \ref{eq-DSD} and \ref{eq-RSD}) can be always applied no matter what value the interval number $n$ is (including $n=1$). Additionally, this "multiple power-law" model can be applied to not only craters but also small bodies such as the main-belt asteroids whose size distribution has also presented power-law breaks (\citealt{Ivezic2001, Parker2008}).

\subsection{Size Distributions of Population 1 and 2 Craters}   \label{sec-Popu}

\begin{figure}
\centering
\includegraphics[width=8cm]{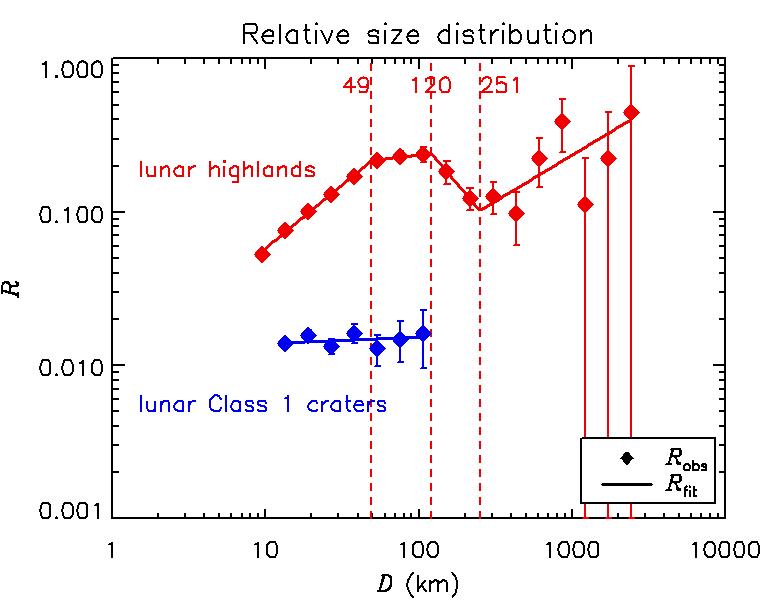}
\caption{Relative size distributions of the lunar highland craters (red) and lunar Class 1 craters (blue), typical of those of Population 1 and 2 craters. For each population, the observed size distribution (\citealt{Strom2005}) is plotted with diamonds, while its best-fit is the curve in the same color. The power-law transitions of fitted Population 1 size distribution are signed with vertical dashed lines in red.}
\label{fig-RSFD_Popu}
\end{figure}

Here the size distributions of two crater populations will be fitted to derive their power-law slopes and interval limits. The observational data are from \cite{Strom2005}. The lunar highland craters and lunar Class 1 craters (fresh craters with  pristine morphologies and well-defined ejecta blankets), are taken as typical of Population 1 and Population 2. It is seen in Figure \ref{fig-RSFD_Popu} that the former has a RSD of complex shape characterized by three transitions over $D$ range about 10--2500\,km, while the latter has a smooth shaped RSD over about 10--100\,km. Therefore, the Population 1 and 2 size distributions are assumed to be four connected power laws and simply a single one, respectively.

Hereafter the parameters and variables involved in the multiple power-law model can have an extra initial subscript $p$ = 1 or 2 referring to Population 1 or Population 2. Equation \ref{eq-RSD} leads to
\begin{equation}    \label{eq-lgRSD}
  \lg R_p = \lg (\alpha_{pi} C_{pi}) + (2-\alpha_{pi}) \lg D_p,   \quad (D_{pi} \le D_p \le D_{p(i+1)})
\end{equation}
where according to Equation \ref{eq-C_i},
\begin{equation}
  \lg (\alpha_{pi} C_{pi}) =
  \begin{cases}
    \lg (\alpha_{p0} C_{p0}), & (i=0) \\
    \lg (\alpha_{p0} C_{p0}) + \sum\limits_{j=1}^{i}(\alpha_{pj}-\alpha_{p(j-1)}) \lg D_{pj}. & (i = 1,2,\dots,n-1)
  \end{cases}
\end{equation}
So taking $\lg D_p$ and $\lg R_p$ as independent and dependent variables, parameters $\alpha_{p(0,1,\dots,n-1)}$ and $C_{p0}$ can be estimated by linear least-squares fitting with $D_{p(0,1,\dots,n)}$ given.

The interval number of Population 2 size distribution is $n_2$ =1, since it is modeled as a single power law, and the interval limits $D_{20}$ and $D_{21}$ are simply defined as the minimum and maximum of variable $D_{2}$. The fit of RSD of lunar Class 1 craters results in $\alpha_{20} = 1.96 \pm 0.14$ and $C_{20} = (6.41 \pm 3.60) \times 10^{-3}$.

The Population 1 interval number is set to $n_1$ = 4. The first and last interval limits $D_{10}$ and $D_{14}$ are also the minimum and maximum of $D_{1}$, but $D_{11}$, $D_{12}$ and $D_{13}$ can not be directly determined. Our solution is to attempt combinations of $D_{11} \in$ ($D_{10}$, 100]\,km, $D_{12} \in$ [50, 300]\,km and $D_{13} \in$ [100, 600]\,km excluding those do not satisfy $D_{11} < D_{12} < D_{13}$ and record every $\chi^2$ (weighted sum of squared errors) so as to find the combination leading to the best fit. Specifically, $D_{1(1,2,3)}$ are all attempted in step of $\lg D_1$ = 0.05 at first and then in a halved step crossing their halved ranges centered at the temporary best-fit-leading values repeatedly, until the step is less than $\lg D_1$ = 0.001 (so that the uncertainties of $D_{1(1,2,3)}$ are only $\sim 0.1$\,km). It turns out only given $D_{11} = 49$\,km, $D_{12} = 120$\,km and $D_{13} = 251$\,km, can the fitted parameters $\alpha_{10} = 1.17 \pm 0.61$, $\alpha_{11} = 1.88 \pm 0.88$, $\alpha_{12} = 3.17 \pm 0.80$, $\alpha_{13} = 1.40 \pm 0.15$ and $C_{10} = (7.29 \pm 15.16) \times 10^{-3}$ give rise to the minimized $\chi^2$. Note the poor statistics of large craters contributes most of the uncertainties. With observational data of $D_1 > D_{13}$ excluded and $n_1$ = 3 assumed, approximately the same optimal values of parameters are still found when the same $D_{1(1,2)}$ are given, but the uncertainties are much smaller: $\alpha_{10} = 1.17 \pm 0.04$, $\alpha_{11} = 1.88 \pm 0.07$, $\alpha_{12} = 3.17 \pm 0.10$ and $C_{10} = (7.29 \pm 1.11) \times 10^{-3}$.

The apparent agreements between the fitted and observed RSDs of every crater population are shown in Figure \ref{fig-RSFD_Popu}. Also, the derived slopes $\alpha_{1(0,1,2,3)}$, $\alpha_{20}$ and transition points $D_{1(1,2,3)}$ are found well consistent with \cite{Strom2015}, who estimated $\alpha_{1} =$ 1.2, 2 and 3 for $D_1 \lesssim$ 50\,km, 50\,km $\lesssim D_1 \lesssim$ 100\,km and 100\,km $\lesssim D_1 \lesssim$ 300\,km in turn and $\alpha_{2} =$ 2 for 0.02\,km $\lesssim D_2 \lesssim$ 100\,km. However, it should be pointed out that whether how many intervals a crater size distribution is divided into or where its transitions roughly are is decided visually. The way it is modeled does not necessarily generate a mathematically best fit. For example, if every data point is taken as a power-law transition, $\chi^2$ = 0 will be certainly obtained but senselessly. So we consider our fitted Population 1 and 2 size distributions to be empirical compromises between fitting preciseness and physical meaning. In addition, we caution about the neglected uncertainties resulting from the slight dependence of observed RSDs on the bin size of $D$.

Theoretically, fits of $\alpha_{1(0,1,2,3)}$ and $\alpha_{20}$ are also valid for the crater size distributions of terrestrial planets, but $D_{1(1,2,3)}$ are not because the crater size is determined by both aspects about the target and the impactor. As \cite{Strom2015} indicated, there is a systematic rightward shift of peak diameters of Population 1 RSDs from the Mars to the Moon and then to the Mercury, which is consistent with the increasing mean impact speeds between them and the asteroids originating from the main belt.

\subsection{Partition of Mixed Lunar Craters}  \label{sec-division}

\begin{figure}
\centering
\includegraphics[height=14.5cm,angle=90]{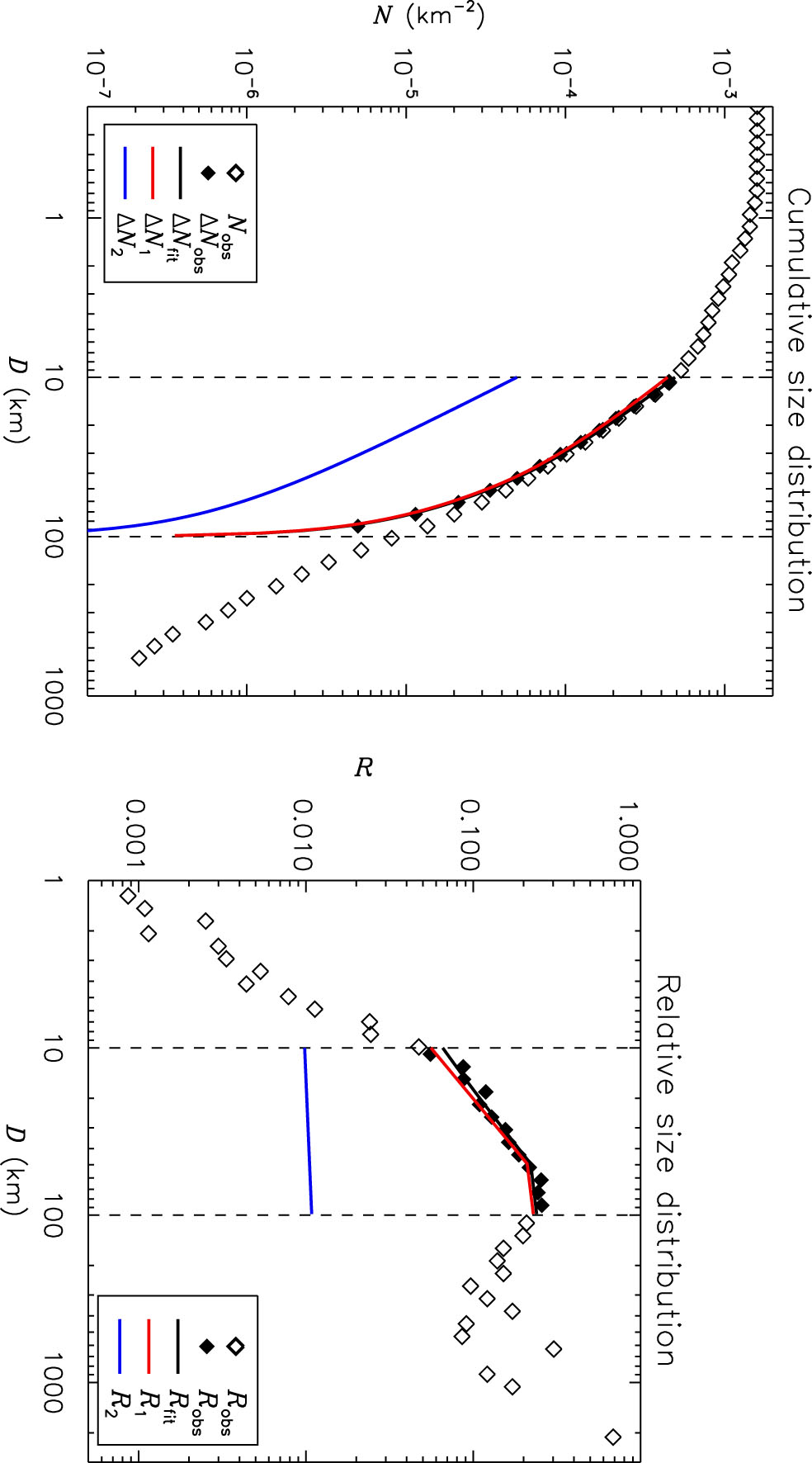}
\caption{Cumulative size distribution (left) and relative size distribution (right) of lunar craters. The size distribution of all the 60645 craters in catalogue LU60645GT (\citealt{Salamuniccar2012}) and the selected 18054 craters with diameters between 10 and 100\,km are plotted with white and black diamonds, respectively. The fit of the latter and its two constituents, size distributions of Population 1 and 2 craters, are plotted with black, red and blue curves in turn. The vertical dashed lines sign the diameter range of selected craters.}
\label{fig-CSD+RSD}
\end{figure}


\begin{figure}
\centering
\includegraphics[width=14.5cm]{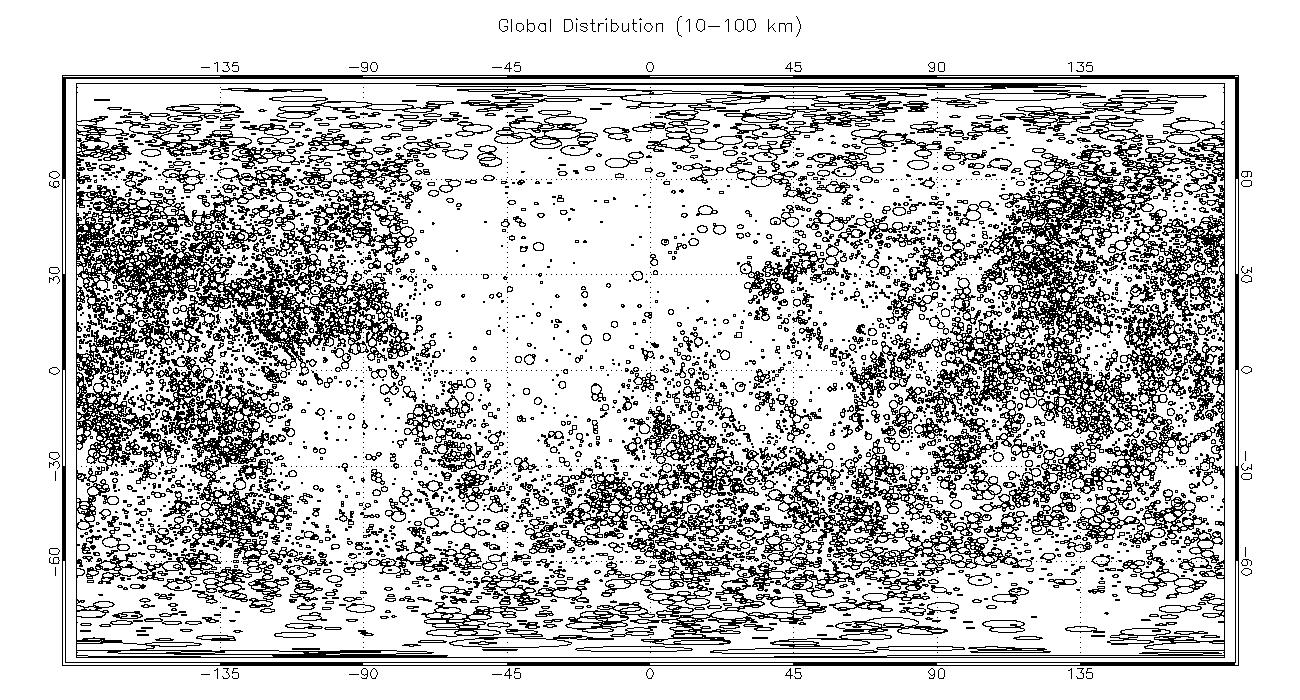}
\caption{Global distribution of lunar craters with diameters between 10 and 100\,km, reproduced with catalogue LU60645GT (\citealt{Salamuniccar2012}). The center of the near side is at (0$\degr$, 0$\degr$). Every crater is considered to be a circle with its rim plotted.}
\label{fig-LU60645GT}
\end{figure}

On the assumption that there have been two impactor populations in the inner solar system, the craters on every unit area of the lunar surface can be taken as a mixture of Population 1 and 2 craters. Ignoring the erosion and saturation, the older the lunar area, the greater the crater density of each population (on condition that this area was formed in the dominant epoch of the relevant impactors). In the $R$ plot, this corresponds to the higher vertical position of the RSD, while the RSD shape of each crater population is invariant. Ignoring the cratering asymmetry (Sect. \ref{sec-asymmetry}), the horizontal position of the RSD is also invariant, regardless of the geographic location of every lunar area. Therefore, each crater population on the whole lunar surface has exactly the same RSD shape as the typical one determined in Section \ref{sec-Popu}, and we can model the size distribution of global lunar craters as a sum of Population 1 and 2 crater size distributions. By fitting the model to the observations, we will see the amounts of two crater populations reveals great disparity.

LU60645GT is a uniform lunar crater catalogue complete up to $\sim D \ge 8$\,km (\citealt{Salamuniccar2012}). The CSD and RSD of all the 60645 craters listed there are shown with white diamonds in Figure \ref{fig-CSD+RSD}. Figure \ref{fig-LU60645GT} shows the craters with diameters from 10 to 100\,km totaling 18054 selected from LU60645GT. The lower limit 10\,km is set to avoid the contamination of secondary craters and to ensure the completeness of this lunar crater sample, while the upper limit 100\,km is set to adapt to the maximum size of observed lunar Population 2 craters (lunar Class 1 craters). The CSD and RSD of this portion of craters are also shown in Figure \ref{fig-CSD+RSD} with black diamonds. The clear similarity of the latter to the Population 1 RSD implies the dominance of Population 1 craters.

Directly applying Equation \ref{eq-CSD2}, the CSDs of Population 1 and 2 craters on the Moon are formulated as
\begin{align}
  N_{1}(D) &=
  \begin{cases}
    C_{10}(D^{-\alpha_{10}}-D_{11}^{-\alpha_{10}}) + C_{11}(D_{11}^{-\alpha_{11}} -100^{-\alpha_{11}}) + N_{1}(100),  &  (10 \le D \le D_{11})  \\
    C_{11}(D^{-\alpha_{11}}-100^{-\alpha_{11}}) + N_{1}(100), & (D_{11} \le D \le 100)
  \end{cases}   \label{eq-N_1}  \\
  N_{2}(D) &= C_{20}(D^{-\alpha_{20}}-100^{-\alpha_{20}}) + N_{2}(100), \qquad  (10 \le D \le 100)    \label{eq-N_2}
\end{align}
where $\alpha_{10}$ = 1.17, $\alpha_{11}$ = 1.88, $\alpha_{20}$ = 1.96 and $D_{11}$ = 49\,km according to Section \ref{sec-Popu} but $C_{10}$, $C_{11}$ and $C_{20}$ now relevant to global lunar craters are not the same as previous values.
Their sum $N(D) = N_{1}(D) + N_{2}(D)$ is the CSD of mixed lunar craters. That unknowns $N_{1}(100)$ and $N_{2}(100)$ consisting in the term $N(100) = N_{1}(100) + N_{2}(100)$ can not be decoupled through fitting is another reason why lunar craters larger than 100\,km are excluded. To describe the size distribution of that 10--100\,km crater sample, $\Delta N(D) = N(D) - N(100)$ is defined and thus
\begin{equation} \label{eq-Delta N}
  \Delta N(D) =
\begin{cases}
    C_{10}(D^{-\alpha_{10}}-D_{11}^{-\alpha_{10}}) + C_{11}(D_{11}^{-\alpha_{11}}-100^{-\alpha_{11}}) + C_{20}(D^{-\alpha_{20}}-100^{-\alpha_{20}}), & (10 \le D \le D_{11})\\
    C_{11}(D^{-\alpha_{11}}-100^{-\alpha_{11}}) + C_{20}(D^{-\alpha_{20}}- 100^{-\alpha_{20}}). & (D_{11} \le D \le 100)
  \end{cases}
\end{equation}

Non-linear least-squares fitting is performed for $\Delta N_{\rm{obs}}$ (Fig. \ref{fig-CSD+RSD}) with Equation \ref{eq-Delta N}. Given $\alpha_{1(0,1)}$, $\alpha_{20}$, $D_{11}$ and dependence of $C_{11}$ on $C_{10}$ (Eq. \ref{eq-C_i}), we derive the best fit $\Delta N_{\rm{fit}}$ together with the estimates $C_{10} = (7.07 \pm 0.01) \times 10^{-3}$ and $C_{20} = (4.55 \pm 0.10) \times 10^{-3}$. Now the comparison between different crater populations mixed in global lunar craters is possible (for 10--100\,km portion). This is what can hardly be done by morphologic classification in observations.

Using estimates of $C_{10}$ and $C_{20}$ and Equation \ref{eq-N_1} and \ref{eq-N_2}, $\Delta N_1(D) = N_1(D) - N_1(100)$ and $\Delta N_2(D) = N_2(D) - N_2(100)$ are obtained. It is seen in Figure \ref{fig-CSD+RSD} that $\Delta N_{1}$ almost completely overlaps $\Delta N_{\rm{fit}}$, indicating the insignificance of Population 2 craters. The nearly undistinguishable divergence is a little more clear between $R_{1}$ and $R_{\rm{fit}}$. (An $R$ plot can illustrate more details than an $N$ plot.) The global densities of 10--100\,km portion of lunar craters are calculated to be $\Delta N_{1}(10) = (43.8 \pm 0.1) \times 10^{-5}$\,km$^{-2}$ and $\Delta N_{2}(10) = (5.0 \pm 0.1) \times 10^{-5}$\,km$^{-2}$ for Population 1 and 2, respectively, i.e., Population 2 craters only make up $(10.2 \pm 0.2) \%$ of this 10--100\,km lunar crater sample. Since Population 2 is deficient in larger craters, its weight in all craters larger than 10\,km should be even less, i.e., $N_{2}(10) / N(10) \lesssim 10 \%$, and it is probably true that $N_{1}(10)$ exceeds $N_{2}(10)$ by more than one order of magnitude (Sect. \ref{sec-intro}).

Additionally, the uncertainties of $\alpha_{10}$, $\alpha_{11}$ (derived with data points of large $D_1$ excluded) and $\alpha_{20}$ are considered. The above results are derived from the optimal values of them determined in Section \ref{sec-Popu}. Eight more cases with each of $\alpha_{10}$, $\alpha_{11}$ and $\alpha_{20}$ added or subtracted by its uncertainty are executed following the same fitting procedure. The maximum weight of Population 2 craters in 10--100\,km sample found when $\alpha_{10}$, $\alpha_{11}$ and $\alpha_{20}$ are all the smallest is $(20.0 \pm 0.2) \%$, while the minimum found when they are all the greatest is $(2.1 \pm 0.2) \%$. So the dominance of Population 1 craters remains.

We caution about the ignorance of the geological variation. The global lunar craters that we take into account can involve the erasing effect, which tends to make the small end of CSD flatter for the older lunar terrains (\citealt{Marchi2009}), and the cratering asymmetry, which can bias the crater size distribution with varying geographic locations. We will discussed the latter influence in Section \ref{sec-asymmetry}.

\section{Discussion}	\label{sec-discussion}

\subsection{Influence of Cratering Asymmetry}   \label{sec-asymmetry}

The cratering asymmetry is the nonuniformity of cratering distribution. A synchronously locked satellite encounters more impactors with larger mean impact speed on its leading side than its trailing side, leading to the leading/trailing asymmetry. Meanwhile, its primary may shields the satellite's near side from impactors or gravitationally focusing the impactors onto it, leading to the near/far asymmetry. In addition, anisotropy of impactors gives rise to the pole/equator asymmetry. Lunar cratering asymmetry has been confirmed both in theory (\citealt{LeFeuvre2008, LeFeuvre2011, Gallant2009, Ito2010, Wang2016}) and in observations (\citealt{Kawamura2011, Oberst2012}).

How does cratering asymmetry influence the crater size distribution? Taking the leading/trailing asymmetry for example, on the one hand, enhancement (diminishment) of impact probability near the lunar apex (antapex) point leads to increase (decrease) of crater number, and on the other hand, enhancement (diminishment) of impact speed there leads to increase (decrease) of crater size. That corresponds to the upward and rightward (downward and leftward) shift of the apex (antapex) RSD in the $R$ plot (assuming impactor size distribution is globally invariant). Apparently, the vertical shift of local RSD alone does not change the RSD shape of global lunar craters, but the horizontal shift does if the size distribution is not a single power law. An RSD involving power-law transitions like that of Population 1 will be extended by horizontal shift and to what degree it is extended is determined by the difference of impact speeds on the apex and the antapex. Note if both the vertical and horizontal shifts exist, the RSD shape will be more twisted than just extended.

Our work has not taken cratering asymmetry into account. If we do, the slope of Population 2 crater size distribution is still valid since it is assumed to be a single power law. However, even the variation of asymmetry degree in the lunar near-side highland (\citealt{Strom2015}) area is neglected, i.e., the fitted RSD of lunar highland craters is exactly a local Population 1 RSD, the global Population 1 RSD can not be directly derived. It should be a sum of every local RSD with varying shifts. The horizontal shifts can be determined by the local geographic positions only if the cratering asymmetry degree generated by Population 1 impactors is known, and the vertical shifts can be obtained only if the ages of geologic units and thus proportion of Population 1 craters obscured by the volcanic resurfacing are known. Both the preconditions are not easy, let alone the latter is dependent on the cratering asymmetry itself (Sect. \ref{sec-chronology}).

Fortunately, the influence of lunar cratering asymmetry on this work seems negligible, because the lunar orbital speed $v_{\rm{M}} \sim$ 1\,km~s$^{-1}$ is much smaller than the encounter speed with the Earth-Moon system of whether Population 1 or 2 impactors $v_{\rm{enc}} \sim$ 20\,km~s$^{-1}$ (\citealt{Gallant2009, Ito2010, LeFeuvre2011, Wang2016}). Applying the Pi-group crater scaling law $D \propto v^{0.44}$ where $v$ is the impact speed (\citealt{Schmidt1987}), the maximum-to-minimum ratio of crater diameter is $\sim [(v_{\rm{enc}} + v_{\rm{M}})/(v_{\rm{enc}} -v_{\rm{M}})]^{0.44}$, equivalent to a negligible $D$ variation of $\sim 5\%$. Alternatively, we can directly adopt the leading/trailing asymmetry amplitude of crater diameter $A_1^D \sim$ 0.02 determined by \cite{Wang2016}, which means the crater diameter near the apex (antapex) is statistically 2\% greater (smaller) than the global average. Taking the average size of lunar highland craters $\sim 30$\,km as that of global Population 1 craters, the rightward shift from antapex RSD to apex RSD is estimated to be $\sim$ 1\,km. The twist of Population 1 RSD is thus negligible and the conclusion in this paper still holds.

\subsection{Revision of Cratering Chronology } \label{sec-chronology}

The cratering chronology method is the technique of age-determination of geologic units on planetary and lunar surfaces by counting craters on them. Its basement is an empirical relationship between geologic age and crater density established using radiometric ages of rock samples from Apollo and Luna missions (e.g., \citealt{Hartmann1981, Neukum1984, Neukum2001}).
The classical cratering chronology has been deduced on the assumption that lunar crater density is globally uniform. That the cratering asymmetry is not considered can lead to overestimate and underestimate of the age where crater density is enhanced and diminished, respectively.

\cite{Morota2005} and \cite{LeFeuvre2011} have revised the cratering chronology using the cratering asymmetry generated by the current impactors. The former estimated the maximum age error due to the cratering asymmetry to be over 20\%. The latter also found that the age error could be $\sim 25\%$ for those geologic units formed in the past $\sim 3.5$\,Gyr, when the impact flux is nearly constant. For the older times, as \cite{LeFeuvre2011} implied, the exponential relationship between the crater density and the geologic age results in a moderate influence of the cratering asymmetry.
However, \cite{Morbidelli2012} provided evidence of a weaker LHB which occurred $\sim 4.1$\,Gya and declined slowly. Given that the Population 1 craters (which are on the oldest regions) make up almost all the lunar craters larger than 10\,km (Sect. \ref{sec-division}), and that their leading/trailing asymmetry degree is likely to be greater than the other population (\citealt{Wang2016}), to include the cratering asymmetry generated by Population 1 impactors in revising the cratering chronology is worth consideration.

\section{Conclusion}	\label{sec-conclusion}

Proportions of Population 1 and 2 craters of the Moon are quantitatively determined. The main results are as follows.

\begin{enumerate}
  \item The multiple power-law model capable of describing a crater size distribution with varying power-law slope is built.
  \item Typical of Population 1 and 2 crater size distributions are fitted, resulting in slopes of the former $\alpha_{10} = 1.17 \pm 0.04$ for $D$ from $\sim$10 to 49\,km, $\alpha_{11} = 1.88 \pm 0.07$ for $D$ from 49 to 120\,km, $\alpha_{12} = 3.17 \pm 0.10$ for $D$ from 120 to 251\,km, $\alpha_{13} = 1.40 \pm 0.15$ for $D$ from 251 to $\sim$2500\,km and single slope of the latter $\alpha_{20} = 1.96 \pm 0.14$ for $D$ from $\sim$10 to $\sim$100\,km.
  \item Size distribution of 10--100\,km lunar crater sample is fitted, leading to the proportion of Population 2 craters in this sample equal to 10\% without uncertainties of $\alpha_{1(0,1)}$ and $\alpha_{20}$ considered, and from 2\% to 20\% with them considered.
\end{enumerate}

Our calculation emphasizes the importance of Population 1 craters and the lunar cratering by their impactors, i.e., the primordial MBAs who dominated during the LHB. The twist of Population 1 crater size distribution due to the cratering asymmetry is noted, but estimated to be too small to influence our conclusion.

\begin{acknowledgements}
We greatly appreciate the information R. Strom gave and the catalogue LU60645GT G. Salamuni{\'c}car provided, and are thankful for the comments by the anonymous reviewer.
This research has been supported by the Key Development Program of Basic Research of China (Nos. 2013CB834900), the National Natural Science Foundations of China (Nos. 11003010 and 11333002), the Strategic Priority Research Program "The Emergence of Cosmological Structures" of the Chinese Academy of Sciences (Grant No. XDB09000000), the Natural Science Foundation for the Youth of Jiangsu Province (No. BK20130547) and the 985 project of Nanjing University and Superiority Discipline Construction Project of Jiangsu Province.
\end{acknowledgements}

\bibliographystyle{raa}
\bibliography{02}

\end{document}